\documentclass[11pt]{article}
\usepackage{latexsym}

\newtheorem{theorem}{Theorem}

\newtheorem{lemma}{Lemma}

\begin{document}

\title{ \bf Boundary Value Problem for $r^2 \,{d^2 f/dr^2}  + f = f^3$ (I): 
Existence and Uniqueness}

\author{Chie  Bing Wang\thanks
  {\small \it Current address:Department of Mathematics, University of California,
  Davis, CA 95616.e-mail:cbwang@math.ucdavis.edu } \\
 \small \it  Department of Mathematics, University of Pittsburgh \\
 \it Pittsburgh, PA 15260.}

\date{}
\maketitle

\begin{abstract}
In this paper we study the equation $r^2 \,{d^2 f/dr^2}  + f = f^3$ 
with the boundary conditions $f(1)=0$, $f(\infty)=1$ and
$f(r) > 0$ for $1<r<\infty$. The existence of the solution 
is proved by using topological shooting argument. And the uniqueness
is proved by variation method. Using the asymptotics of 
$f(r)$ as $r \to 1$, in the following papers 
we will discuss the global solution for $0<r<\infty$,
and give explicit  asymptotics of $f(r)$ as $r \to 0$
and as $r \to \infty$, and the connection formulas for the parameters
in the asymptotics. 
Based on these results, we will solve the boundary value problem
$f(0) =0$, $f(\infty) =1$, which is the goal of this work.
Once people discuss the regular solution of this
equation, this boundary value problem must be considered.
 This problem is useful to study
 the Yang-Mills coupled equations, 
and the method used for this equation is applicible to 
other similar  equations.
\end{abstract}



\markboth{C. B. Wang}{ Monopole in the Pure SU(2) Gauge Theory}

\setcounter{equation}{0}
\section{Introduction}

 Consider the following
 boundary value problem
    \begin{eqnarray}
  & &    r^2 f'' + f = f^3, 0<r<\infty,    \label{eq1.1}   \\
 & & f(r)  \to 0,  \,  {\rm as \,\,} r \to 0,  \label{eq1.2}  \\
 & & f(\infty)=1,  \label{eq1.3}
    \end{eqnarray}
where $'$ means $d/dr$.
This problem was proposed in \cite{wu}(Wu and Yang) for studying the 
monopole solution in the pure SU(2) gauge field theory. 
The solution to this problem is usually called Wu-Yang solution.
And when people study the Yang-Mills coupled equations, for example, in 
 \cite{breitenlohner}  \cite{hastings} \cite{smoller},
this equation is always  considered.
The regular solution to this equation only comes out from this 
boundary value problem or equivalently $f(\infty)=-1$.
So it is useful to give a complete study for the existence, uniqueness,
asymptotics and connection formulas for the
parameters in the asymptotic formulas. 
The readers who are interested in the physics background are
refered to \cite{actor} \cite{malec}. 

Wu and Yang \cite{wu}, Protogenov \cite{protogenov}
and Breitenlohner, Forg\'acs and Maison \cite{breitenlohner}
obtained that the solution to this boundary value problem
has  the asymptotics 
\[
 f(r) \sim \alpha \, r^{1\over2} \sin \left( {\sqrt{3} \over 2} \log r + \beta \right),
\]
as $r \to 0$, and
\[
  f(r) \sim 1 + { \gamma \over r},
\]
as $r \to \infty$ for some parameters $\alpha, \beta, \gamma$.
The current work is motivated to find the formulas for the parameters
$\alpha, \beta$ and $\gamma$, which are called connection formulas
for this problem.

  In this paper and in \cite{wang1} \cite{wang2}, we study this
 boundary value problem and finally give the connection formulas.
We will show that any solution to this problem has infinitely many zeros, 
and the zeros have upper bound. So the largest zero $r=r_0$ exists.
Since the equation is invarient under the scaling transformation
$r \to c \,r$, we just need to discuss $r_0 = 1$. 
To study the boundary value problem, we first consider 
the existence and uniqueness of another
boundary value problem $f(1)=0, f(\infty) =1$, and $f(r) >0$ for
$r>1$, which is the work of this paper. By using shooting argument and 
variation method, we prove that 
this problem has a unique solution, and the solution has
asymptotics
\[
  f(r) \sim a^* \, \log r,
\]
as $ r \to 1$, for a positive constant $a^*$.

 In \cite{wang1} we will find exact formula for this number $a^*$ by using
analytic continuity method to study the analytic property of the solution
at $r=1$. In \cite{wang2}, we will discuss the global solution to the
boundary value problem (\ref{eq1.1}) (\ref{eq1.2})  and (\ref{eq1.3}),
and give the asymptotics and connection formulas. 
The method used in these papers would be applicible to study a more general equation
$r^2 f^{''} = F(f)$, where $F(f)$ is a polynomial of $f$.

  To study  equation (\ref{eq1.1})    we put
   \begin{equation}
      r= e^x,   f(r) = y(x).  \label{eq1.15}
   \end{equation}
Then (\ref{eq1.1}) is   changed to 
 \begin{equation}
   y'' -y' + y = y^3.   \label{eq1.20}  
 \end{equation}
In Sect. 2 we prove that there is a solution $y^*(x)$ to equation 
(\ref{eq1.20}) for $x >0$, such that $y^*(0)=0, y^*(\infty)=1$
and $y^*(x) >0, x >0$. The method we use in this paper is 
the one dimensional shooting argument which has been widely used
to discuss boundary value problems. In Sect. 3 we show the solution 
$y^*(x) (x >0)$ is unique and strictly monotone by 
using variation method. 
In the next paper \cite{wang1}, we will discuss the number $a^* = y^{*^{'}}(0)$,
which will be used to analyze the global solution $(-\infty < x < \infty)$
\cite{wang2}.

\setcounter{equation}{0}
\section{Existence of the Solution}
   Let us consider the following problem
   \begin{eqnarray}
    & & y'' - y'= y^3 - y, \,\, 0<x<\infty, \label{eq2.1}  \\
    & & y(0) = 0, y'(0) = a.   \label{eq2.2}
   \end{eqnarray}
In this section we show that there is a positive value of $a$,
such that the solution to  (\ref{eq2.1}) and (\ref{eq2.2})
satisfies
    \begin{eqnarray*}
    & & y(\infty)=1,   \\
    & & y'(x) > 0, \,\, 0<x<\infty.
    \end{eqnarray*}
First let us state a basic fact in ordinary differential
equation theory \cite{coddington}.
\begin{lemma} \label{Lemma2.1}
For any $a$, there is a unique bounded solution $y(x, a)$ to
(\ref{eq2.1}) (\ref{eq2.2})   in a neighborhood of $0$. 
Specially when $a=0$, $y \equiv 0$.
\end{lemma}

   We then analyze the behaviour of the solution when
$a$ is large or small. It will be shown below that
when $a$ is large, $y$ crosses $1$ before $y'$ crosses $0$, and
when $a$ is small, $y'$ crosses $0$ before $y$ crosses 1.
Then we show there is a value of $a$, such that $y(x,a)$ does not
cross $1$, and $y'(x,a)$ does not cross $0$, and $y(\infty, a)=1$.
This is the so called shooting method.

\begin{lemma}  \label{Lemma2.2}
When $a>{1 \over \sqrt{2}}$,  the solution $y(x)$ to
(\ref{eq2.1}) (\ref{eq2.2}) satisfies
   \begin{eqnarray}
   & & y(x^+) > 1,   \\
   & & y'(x) > 0, \,\, 0 \le x \le x^+,
   \end{eqnarray}
where $x^+ = (a^2 - 1/2)^{-1/2} + 1$.
\end{lemma}

\noindent {\it Proof.} Let
   $$
   v(x) = 1 - y(x).
   $$
Then (\ref{eq2.1}), (\ref{eq2.2}) become
    \begin{eqnarray}
   & & v'' - v' = 2 v - 3 v^2 + v^3,  \label{eq2.5} \\
   & & v(0) = 1, v'(0) = -a.     \label{eq2.6}
     \end{eqnarray}
Multiplying equation  (\ref{eq2.5} ) by $v'$ and integrating, we arrive at
   $$ v'^2 = 2 v^2(1- {1 \over 2} v)^2+ (a^2 - {1 \over 2} )
        + 2 \int_0^x \, (v'(s))^2 \, ds,
   $$
for $x > 0$.
When $a > {1 \over \sqrt{2}}$, the right hand side of the
above equation is always positive. And because $v'(0)=-a<0$,
we have
   \begin{eqnarray*}
  v'(x) &=& - \left(  2 v^2 (1- {1 \over 2} v)^2 + (a^2 - {1 \over 2} )
           + \int_0^{x} \, {v'(s)}^2 \, ds \right)^{1/2}  \\
        &<& - \left( a^2 - {1 \over 2} \right)^{1/2}.
   \end{eqnarray*}
Hence
   \begin{eqnarray*}
    v(x^+) &=& v(0)+ \int_0^{x^+} \, v'(s) \, ds  \\
         &<& 1 - \int_0^{x^+} \sqrt{a^2 - {1 \over 2}} \, ds \\
         &=& - \sqrt{a^2 - {1 \over 2}} <0,  \\
    v'(x) &<& 0,  \,\,\, 0 \le x \le x^+,
   \end{eqnarray*}
or equivalently
   \begin{eqnarray*}
    y'(x^+) &>& 1,  \\
    y'(x) &>& 0, \,\,\,  0 \le x \le x^+.
   \end{eqnarray*}
So the lemma is proved.   \,\,\,\, $\Box$

\begin{lemma} \label{Lemma2.3}
There is $a^- > 0$, such that if $a \in (0,a^- ]$, the solution
$y(x)=y(x,a)$    satisfies
  \begin{eqnarray}
   y'(x^-) &<& 0,  \\
   y(x) &>& 0, 0<x \le x^-,
  \end{eqnarray}
where
   $ x^- = { 5 \pi \over 3 \sqrt{3}}. $
\end{lemma}

\noindent {\it Proof.} Let 
   \begin{equation} 
    y(x) = a \, \,w(x).  \label{eq2.15}
   \end{equation}
Then  (\ref{eq2.1}), (\ref{eq2.2}) become
   \begin{eqnarray}
     w'' - w' + w = a^2 w^3,     \\
     w(0)=0, w'(0) = 1.
   \end{eqnarray}
As $a \to 0$, $w(x)$ uniformly on compact intervals in $x$ tends to
the solution of the problem
    \begin{eqnarray}
     W'' - W' + W = 0,     \\
     W(0)=0, W'(0) = 1.
   \end{eqnarray}
It's not hard to see that the solution of this problem is
    \[ W(x) = {2 \over \sqrt{3}} 
      e^{ {1 \over 2} x} \sin \left( {\sqrt{3} \over 2} x \right).
    \]
We see that
   \begin{eqnarray}
    W'(x^-) &<& 0,     \\
    W(x) &>& 0,  0<x \le x^-.
   \end{eqnarray}
Thus there exists $a^- > 0$, such that if $a \in (0,a^- ]$, there is
  \begin{eqnarray}
   w'(x^-) &<& 0,  \\
   w(x) &>& 0, 0<x \le x^-.
  \end{eqnarray}
By equation  (\ref{eq2.15}), the lemma is proved.  \,\,\,\, $\Box$

   For the solution $y(x,a)$ of (\ref{eq2.1}), (\ref{eq2.2}), 
where again $a = y'(0)$, let us define 
   \begin{eqnarray}
  S^+ &=& \{ a>0 |\, y {\rm  \,\,crosses \,\,1\,\, 
            before\,\,} y' {\rm \,\, crosses \,\,0} \}, 
           \label{eq2.18}   \\
  S^- &=& \{ a>0 |\, y'{\rm  \,\,crosses\,\, 0\,\, 
            before\,\,} y {\rm \,\, crosses\,\, 1} \}, 
           \label{eq2.19}   
   \end{eqnarray}
Lemma 2 shows that $( 1/ \sqrt{2}, \infty) \subset S^+$, and
Lemma 3 shows that $( 0, a^-) \subset S^-$.

\begin{theorem} \label{Theorem2.4}
There is a solution to the following problem
   \begin{eqnarray}
     & & y'' - y'= y^3 - y, \,\, 0<x<\infty, \label{eq2.20}  \\
 (P^+ )   & & y(0) = 0, y(\infty) = 1,   \label{eq2.21}  \\
     & & y(x) > 0, \,\, 0< x < \infty.   \label{eq2.22}
   \end{eqnarray}
\end{theorem}

\noindent {\it Proof.} By Lemma 2 and Lemma 3, we have that 
$S^+$ and $S^-$ are non-empty
sets. By the definition of $S^-$ and $S^+$, we see that
they are disjoint  sets. By implicit function theorem, it is not difficult
to show that $S^+$ and $S^-$ are open sets. Thus
  \[ (0, \infty) \setminus ( S^- \cup S^+ ) \ne \emptyset. \]
Hence there is $a^* >0, \, a^* \notin S^- \cup S^+ $,
such that $y(x, a^*)$ satisfies
   \begin{eqnarray}
     y'(x,a^*) >0, 0< x < \infty,   \\
     y(x,a^*) < 1, 0<x < \infty.
   \end{eqnarray}
So $y(\infty, a^*) = b$, where $0<b \le 1$. 
  
    Let us show $b=1$.  There exists $x_0 >0$ such that
when $x_0 < x < \infty$,
    \[  {b \over 2} < y(x) < b.  \]
If $b <1$, we have from (\ref{eq2.20})
   \begin{eqnarray*}
   y'(x) &=& e^x \int_{x_0}^{x} e^{-s} y(s) ( y^2(s) - 1) \, ds \\
         &\le& e^x \int_{x_0}^{x} e^{-s}
                 {b \over 2} ( b^2 - 1) \, ds    \\
         &=& { b(b^2-1) \over 2} (e^{x-x_0} -1) \to - \infty,
   \end{eqnarray*}
as $x \to \infty$,which is a contradiction, since $y(\infty, a^*)$ exists. 
So $b=1$. \,\,\,\, $\Box$

\setcounter{equation}{0}
\section{Uniqueness of the Solution}

   In the last section we have proved that problem $(P^+ )$ has a solution.
Now, we show that the solution is also unique. Let us start from
the following lemma.

\begin{lemma} \label{Lemma3.1}
If $y(x)$ is a solution to the problem  $(P^+)$
   \begin{eqnarray}
   & &  y'' - y' + y=y^3, \,\, 0<x< \infty,  \label{eq3.1}   \\
   & &  y(0) = 0, y(\infty) = 1,  \label{eq3.2}  \\
   & &  y(x) > 0, \,\,\, 0<x< \infty,  \label{eq3.3}
   \end{eqnarray}
then $y(x)$ has the following properties
   
   (i) \, \, \,  $ y(x) < 1, 0 \le x < \infty$.

   (ii) \, \, $ y'(x) > 0, 0 \le x< \infty$.  And $y'(\infty) =0$.

   (iii) \,\, $y(x) = 1 - c \, e^{-x} + O(e^{-2x})$,
    \,\, $y'(x) = c \, e^{-x} + O( e^{-2 x} )$, and then
    $ { y'(x) \over y(x) - 1 } = -1 + O(e^{-x})$,
    as $x \to \infty$, where $c >0$.
\end{lemma}

\noindent {\it Proof.}
  (i)  Suppose $x_1 >0$ is the first point, such that $y(x_1)=1$. 
By the uniqueness of the solution, $y'(x_1) > 0$. We then claim that
$y'(x) > 0$, for $x_1<x<\infty$. If not, suppose $x_2 > x_1$ is the 
first point such that $y'(x_2) = 0$. Then since $y'(x) > 0$,
for $x_1 < x < x_2$, there is
   \begin{eqnarray*}
     y'' &=& y' + y ( y^2 -1) > 0,  \\
     y'(x_2) &=& y'(x_1) + \int_{x_1}^{x_2} y''(s) \, ds  \\
             &>& y'(x_1) > 0,
   \end{eqnarray*}
which is a contradiction. So $y'(x) > 0$, for $x_1 <x< \infty$.
Then we can not get $y(\infty) = 1$, which is a contradiction.
Thus $y<1$.

  (ii) If $y'(0) = 0$, then since $y(0) = 0$, by the uniqueness 
of the solution, we see that $y \equiv 0$, which is a contradiction.
So $y'(0) > 0$. 
   
   Now suppose $x_3 > 0$ is the first point, such that $y'(x_3) =0$.
By equation  (\ref{eq3.1}) and (i), there is
 \[ y'(x)=e^x \int_{x_3}^{x} e^{-s} y(s) ( y^2(s) - 1) \, ds <0, \]
for $x > x_3$. Then since $y(x_3) < 1$, we can not have $y(\infty) =1$.
This is a contradiction. So $y'>0$, for all $x \ge 0$.

     If we do not have $y'(\infty) =0$, then there are  (small)
$\epsilon >0$,(large) $ x_0 >0$, such that $y'(x_0) > \epsilon$, and
$y(x) - y^3(x) < \epsilon$ for $x \ge x_0$. By equation  (\ref{eq3.1})
we have
    \[  y''(x_0) = y'(x_0) - (y(x_0)-y^3(x_0) ) >0, \]
which implies that $ y''(x) >0$ in a neighborhood of $x_0$. So
$y'(x)$ is increasing in this neighborhood. By equation  (\ref{eq3.1})
   \begin{eqnarray*}
     y''(x) &=& y'(x) - (y(x)-y^3(x) )   \\
            &>& y'(x) - \epsilon \ge y'(x_0) - \epsilon > 0,
   \end{eqnarray*}
when $x \ge x_0$.
We see that $y''(x)$ keeps positive, and $y'(x)$ keeps
increasing for $x \ge x_0$, which is a contradiction since
$y(\infty)=1$. Therefore we have $y'(\infty) =0$. 

   (iii)
Let $y_1=y, y_2=y'$, and change equation  (\ref{eq3.1}) into the system
        \begin{eqnarray*}
        y_1' &=& y_2,   \\
        y_2' &=& y_2 - y_1 +y_1^3.
        \end{eqnarray*}
It is easy to see that $(1,0)$ is a saddle point in the phase
plane. Since $(y_1(\infty), y_2(\infty))= (1,0) $,
by the stable manifold theorem (\cite{coddington} \cite{perko}) we get
that as $x \to \infty, \, \,(y_1(x), y_2(x))$ lies on the stable
manifold. And by standard argument \cite{perko} we have
     \begin{eqnarray*}
      y(x) &=& 1 - c \, e^{-x} + O(e^{-2x}),   \\
      y'(x) &=& d \, e^{-x} + O ( e^{-2 x} ),
     \end{eqnarray*}
as $x \to \infty$, for some constants $c,d$.
Because $y(x) < 1$ for $x > 0$, $c$  can not be negative.
If $c=0$, we convert the equation
(\ref{eq3.1}) into integral equation by Green function.
By contraction argument we get $ y \equiv 1$, which is a
contradiction. Thus $c >0$. By
      \[ y(x) = 1 - \int_x^{\infty} y'(s) \, ds ,  \]
we see that $c=d$.  \,\,\,\, $\Box$

   To prove the uniqueness, we use variational method.
Suppose $y(x,a_1)$ is a solution to $(P^+)$.
We will show  in this section that when $a$ increases
a little from $a_1$, $y(x,a)$ crosses $1$ at some point.
We then show the root $x$ of $y(x,a)=1$ is moving left
while $a$ is increasing further more, which means
for any $a > a_1$, $y(x,a)$ does not satisfies $y(\infty,a)=1$,
i.e. they are not solutions to $(P^+)$. So we can show the
solution is unique. 

    For simplicity we do not directly discuss $(P^+)$. Instead
we consider the original equation (\ref{eq1.1}).
Suppose $f(r,a)$ is a solution to the following problem
 \begin{eqnarray}
 & & r^2 f'' + f = f^3, \,\, 1<r<\infty,    \label{eq3.31}   \\
 & & f(1) = 0,  \,  f'(1)=a,  \label{eq3.32}    
 \end{eqnarray}
where $a>0$.  It is easy to see that (\ref{eq3.31}) (\ref{eq3.32})
are equivalent to (\ref{eq2.1})  (\ref{eq2.2}).
Define
    \begin{equation}
      \psi(r,a) = {\partial f(r,a) \over \partial a}.  \label{eq3.33}
    \end{equation} 

\begin{lemma}   \label{Lemma3.2}
(i) If $f(r,a)$ crosses $1$ at a point $r=r_1 >1$, and 
$f(r,a) >0$ for $1 < r \le r_1$, then there is
   \begin{equation}
     \psi(r_1, a) > 0,  \label{eq3.35}
   \end{equation}
for $1 < r \le r_1$.

(ii)  If for some $a=a_1$, $f(r,a_1)$  satisfies $f(r,a_1)>0$, for $r >1$,
and  $f(\infty,a_1)=1$, then
    \begin{equation}
     \psi(r, a_1) > 0,  \psi'(r,a_1) >0,  \label{eq3.36}
   \end{equation}
for  $r > 1$.
\end{lemma}

\noindent {\it Proof.}
(i) By the definition of $\psi$ (\ref{eq3.33}), $\psi$ satisfies
     \begin{eqnarray}
      & & r^2 \psi'' + \psi = 3 f^2 \psi,  \label{eq3.41}  \\
      & & \psi(1) = 0, \psi'(1) = 1.   \label{eq3.42}
     \end{eqnarray}
By (\ref{eq3.31}) and (\ref{eq3.41}), we have
     \begin{equation}
        r^2 ( f' \psi - f \psi' )' = -2 f^3 \psi.  \label{eq3.44}
     \end{equation}

    Assume for contradiction $r_0 \in (1, r_1]$ is the
first point, such that $\psi(r_0,a) =0$. 
Then since $r_0$ is the first zero of $\psi(r, a)$ after $r=1$,
and $\psi(1,a)=1>0$, we have $\psi^{'} (r_0, a) \le 0$.
If $\psi^{'} (r_0, a) = 0$, 
by the uniqueness of
solution, $\psi(r,a) \equiv 0$, which is a contradiction.
So $\psi^{'}(r_0, a) < 0$. Then since $f(1)=\psi(1)=0$,
we get from (\ref{eq3.44})
    \[  f'(r_0) \psi(r_0) - f(r_0) \psi'(r_0) = 
        -2 \int_1^{r_0} {f^3(s) \psi(s) \over s^2} \, ds <0.  \]
By the assumption $\psi(r_0)=0$, we arrive at
    \[ - f(r_0) \psi'(r_0) < 0. \]
Then $f(r_0)>0$ implies that $\psi'(r_0) >0$, which is a contradiction.
So (\ref{eq3.35}) is true.

(ii) Recalling the relation between $y(x,a)$ and $f(r,a)$ (\ref{eq1.15}), 
and  by Lemma 4 (i) , we have $0 < f(r,a_1) <1$ for $r >1$,
which implies $f(r,a_1)$ has no singularity in $(1, \infty)$.
Then by the same argument as above, we have $\psi(r, a_1) > 0$
for $r >1$.  Now suppose $r_2 >1$ is the first point such that
$\psi'(r_2,a_1) =0$, then we have
    \[  f'(r_2) \psi(r_2) = f(r_2) \psi'(r_2)  
        -2 \int_1^{r_2} {f^3(s) \psi(s) \over s^2} \, ds <0.  \]
This is a contradiction because $\psi(r_2)>0$, and $f'(r_2)>0$
by Lemma 4 (ii). So the lemma is proved. \,\,\,\, $\Box$

\begin{lemma}  \label{Lemma3.3}
If  $f(r,a_1)$ is a solution to (\ref{eq3.31}), (\ref{eq3.32}),
satisfying   $f(\infty,a_1)=1$ and $f(r, a_1)>0$ for $r>1$,  
then there exists $\bar{\epsilon} >0$, such that
for any $\epsilon \in (0,\bar{\epsilon}]$, $f(r,a_1+ \epsilon)$
crosses $1$ at some point $r_0 > 1$, and $f(r,a_1+\epsilon) >0$
for $1 < r \le r_0$.
\end{lemma}

\noindent {\it Proof.}
By Lemma 5 (ii) and Lemma 4 (ii), there exist 
$\bar{\epsilon} >0, r_2 > r_1 >1$, such that
     \begin{equation}
        f(r,a) > f(r, a_1), f'(r,a) > f'(r,a_1), \label{eq3.61}
     \end{equation}
for $1 < r \le r_2, a_1 < a \le a_1 + \bar{\epsilon}$, and
     \begin{equation}
         f(r,a_1) > {1 \over \sqrt{2}},  \label{eq3.62}
     \end{equation}
for $r \ge r_1$. Let $v(r,a) = f(r,a) - f(r,a_1)$ for 
$a_1 < a \le a_1 + \bar{\epsilon}$. Then $v$ satisfies
the equation
     \begin{equation}
      r^2 v'' = (f^2 + f f_1 + f_1^2 -1 ) v, \label{eq3.63} 
     \end{equation}
where $r >1$.  When $r_1 \le r \le r_2, a_1 < a \le a_1 + \bar{\epsilon}$, 
by (\ref{eq3.61}), (\ref{eq3.62}) and (\ref{eq3.63})
we have
    \begin{equation}
     v(r,a) > 0, \,\, v'(r,a)>0, \,\, v''(r,a) >0.  \label{eq3.65}
    \end{equation}
By  (\ref{eq3.62}) and (\ref{eq3.63}), 
we see that (\ref{eq3.65})
is true for all $r \ge r_1$. Therefore we have
      \[ f(r,a) > f(r,a_1) + ( f(r_1,a) - f(r_1, a_1) )  \]
for all $r \ge r_1$.  Since $f(\infty,a_1) = 1, f(r_1,a)-f(r_1,a_1)>0$,
we then conclude that $f(r,a)$ crosses $1$ at some point for all
$ a \in (a_1, a_1+\bar{\epsilon}]$, and $f(r,a)$ keeps positive and
finite before it crosses $1$. \,\,\,\, $\Box$

\begin{theorem}  \label{Theorem3.4}
Suppose $f(r,a)$ is a solution to (\ref{eq3.31}), (\ref{eq3.32}).
There is a unique value $a=a^*$,  such that $f(\infty,a^*) =1$,
and  $f(r,a)>0$ for $r >1$.
Thus the problem $(P^+)$ has a unique solution $y^*(x)=y(x,a^*)$,
and $y^{*}(x)$ has the asymptotics
\[
  y^*(x) \sim a^* x,
\]
as $x \to 0$, and
\[
  y^*(x) = 1 - c \, e^{-x} + O(e^{-2 x}),
\]
as $x \to +\infty$.
\end{theorem}

\noindent {\it Proof.}
Theorem 1  has shown that such value of $a$ exists. Now suppose
$a_1 >0$ is a value of $a$, such that $y(x,a_1)$ solves $(P^+)$.
We want to show for any $a>a_1$, $y(x,a)$ does not satisfies
$(P^+)$.

    Consider the equation (\ref{eq3.31}). Set
    \[ D = \{ a>a_1 | f(r,a)=1 {\rm \,\,for\,\,some\,\,}r>1 \}.  \]
By Lemma 6, $(a_1, a_1+\bar{\epsilon}] \subset D$.  Let
$r_1=r_1(a)>1$ be the least root of $f(r,a)=1$, for $a \in D$.
By  implicit function theorem, $r_1$ is a
differentiable function of $a$ on $D$ and
  \[ f'(r_1, a) {d r_1(a) \over d a} + \psi(r_1,a) = 0.  \]
By Lemma 6, the conditions in Lemma 5 (i) are satisfied for
$a = a_1 + \bar{\epsilon}$ . So
$ \psi(r_1,a_1 + \bar{\epsilon}) > 0$. 
Since $f'(r_1,a_1 + \bar{\epsilon})>0$ 
($f(r,a_1 + \bar{\epsilon})$ crosses 1 at $r_1$) 
it follows  that
  \begin{equation}
   {d r_1(a) \over d a} <0,    \label{eq3.100}
  \end{equation}
for $a = a_1 + \bar{\epsilon}$ . Lemma 5 (i) implies 
(\ref{eq3.100}) is true for all $ a > a_1 + \bar{\epsilon}$.
Therefore as $a$ increases $r_1(a)$ monotonically decreases.
Thus $D=(a_1, \infty)$, which means for $a>a_1$ \,
$y(x,a)$ does not satisfies $(P^+)$ by Lemma 4. If there
is another value of $a$ which is less than $a_1$,
such that $y(x,a)$ satisfies $(P^+)$,
then by the above argument, $y(x,a_1)$ does not satisfies $(P^+)$,
which is a contradiction.
So we have proved the theorem.   \,\,\,\, $\Box$

\section*{Acknowledgments}
The author sincerely thanks Professor J. B. McLeod 
and  Professor W. C. Troy for 
 helpful discussions.



\begin{thebibliography}{99} 
\bibitem{actor}
  Actor, A.: {\em Classical solutions of SU(2) Yang-Mills theories },
  Rev. Mod. Phys. {\bf 51}(1979), 461-525  .

\bibitem{breitenlohner}
  Breitenlohner,P., Forg$\grave{a}$cs, P., Maison, D.:
  {\em Static spherically symmetric solution of
   the Einstein-Yang-Mills equations},
	 Commun. Math. Phys. {\bf 163}, 141-172(1994).


\bibitem{coddington}
  Coddington, E.A., N. Levinson, N.:
   {\em  Theory of Ordinary Differential Equations},
     McGraw-Hill,New York, 1955.



\bibitem{hastings}
  Hastings, S.P., McLeod, J.B.,  Troy, W.C.:
  {\em Static spherically symmetric solutions of a
    Yang-Mills field coupled to a dilation},
  Proc. R. Soc. Lond. A {\bf 449}, 479-491(1995).


\bibitem{malec}
  Malec, E.: {\it On classical solutions of nonabelian gauge theories},
  Acta Phys. Polonica B {\bf 18} (1987), 1017-1064.

\bibitem{perko}
  Perko, L.:
  {\em Differential Equations and Dynamical Systems},
     2nd ed. Springer-Verlag, New York, 1996.



\bibitem{protogenov}
  Protogenov, A.P.:
  {\em Bag and multimeron solution of the classical
   Yang-Mills equation},
  Phys. Lett. B {\bf 87}, 80-82(1979).

\bibitem{smoller}
  Smoller, J.A., Wasserman, A.G., Yau, S.T.,
	McLeod, J.B.:
  {\em Smooth static solutions of the
  Einstein/Yang-Mills equations},
  Commun. Math. Phys. {\bf 143}, 115-147(1991).

\bibitem{wang1}
  Wang, C. B.: {\em Boundary value problem for $r^2 \,{d^2 f/dr^2}  + f = f^3$(II):
  connection formula}, preprint.

\bibitem{wang2}
  Wang, C. B.: {\em Boundary Value Problem for $r^2 \,{d^2 f/dr^2}  + f = f^3$(III):
   global solution and asymptotics}, preprint.

\bibitem{wu}
  Wu, T.T,  Yang, C.N.:
  {\em Some solutions of the classical isotopic gauge
   field equations},
  in {\it Properties of
  Matter Under Unusual Conditions}, edited by
  H. Mark and S. Fernbach,
  pp. 349-354. Interscience,NewYork, 1969.


\end{thebibliography}
\end{document}